\documentclass[final]{jfm}

\usepackage{graphicx}
\usepackage{newtxtext}
\usepackage{newtxmath}
\usepackage{xcolor}
\usepackage{natbib}
\usepackage{hyperref}
\hypersetup{
    colorlinks = true,
    urlcolor   = blue,
    citecolor  = black,
}

\newcommand{\RomanNumeralCaps}[1]
\linenumbers

\title{On the dynamics of a collapsing bubble in contact with a rigid wall}

\author{Mandeep Saini\aff{1}\corresp{mandeep.saini@sorbonne-universite.fr},
  Erwan Tanne\aff{2},
  Michel Arrigoni \aff{2},
  Stephane Zaleski \aff{1,3}
 \and Daniel Fuster\aff{1}  \corresp{\email{daniel.fuster@sorbonne-universite.fr}}}

\affiliation{\aff{1}Sorbonne Universit\'e and CNRS UMR 7190, Institut Jean le Rond $\partial$'Alembert F75005 Paris, France
\aff{2}ENSTA Bertagne UMR 6027 - IRDL F-29806, Brest, France
\aff{3}Institut universitaire de France, UMR 7190, Institut Jean le Rond $\partial$'Alembert F75005 Paris, France}

\begin{document}
\maketitle

\begin{abstract}
This work reveals that  the dynamic response of a spherical cap bubble in contact with a rigid wall depends on the effective contact angle at the instant prior to collapse. This parameter allows us to discriminate between two regimes in which the mechanisms of interaction between the collapsing bubble and its surrounding medium differ significantly: When the contact angle is smaller than 90 degrees a classical jet directed towards the wall is observed whereas if the initial contact angle is larger than 90 degrees an annular re-entrant jet  parallel to the wall appears. We show that this change of the behaviour can be explained using the impulse potential flow theory for small times which shows the presence of a singularity on the initial acceleration of the contact line when the contact angle is larger then 90 degrees.  Direct Numerical Simulations show that although viscosity regularises the solution at $t > 0$, the solution remains singular at $t=0$. In these circumstances numerical and experimental results show that the collapse of flat bubbles can eventually lead to the formation of a vortex ring that unexpectedly induces long-range effects. The role of the bubble geometry at the instant of maximum expansion on the overall collapse process is shown to be well captured by the impulse potential flow theory, which can be easily generalised to other bubble shapes. These results may find direct application in the interpretation of geophysical flows as well as the control and design of bio-medical, naval, manufacturing and sonochemistry applications.
\end{abstract}



\section{Introduction}
\label{sec:intro}
Cavitation, the process of bubble formation and collapse, is a key physics problem because of its applications in nature, mechanics, biomedical, and many other fields \citep{cole1948,lohse2001,ohl2006,ohl2006sonoporation,maxwell2013,fuster2019review}.  In recent studies, \citet{hou2021study} show that cavitation bubbles can enhance the efficiency of solar absorption refrigeration systems, \citet{bhat2021cavitation} and \citet{prado2022review} reviewed the effect of cavitation in wastewater and biomass treatment. \citet{lyons2019infrasound} compared the low frequency sound signals from volcanic eruptions to that of bubble collapse and suspect the bubbles of the order of 100 m are created by the interaction of magma with water.
Bubbles can generally appear from the nuclei sitting at walls \citep{crum1979tensile,atchley1989crevice,borkent2009nucleation}, 
making the interaction with the wall an essential aspect of bubble collapse dynamics.
The wall modifies the symmetry of the pressure field near the bubble that causes unequal interface acceleration and thus high-speed liquid jets leading to a complex cavitation phenomena \citep{popinet2002bubble, brenner2002single,supponen2017collapse}. These jets are important to understand the destructive potential of the cavitation bubbles for several biomedical applications such as the disruption of biological membranes in High Intensity Focussed Ultrasound (HIFU), lithotripsy, histotripsy  and sonoporation techniques \citep{prentice2005membrane,ohl2006sonoporation,maxwell2011cavitation,pishchalnikov2019high}. Other applications concerning erosion, surface cleaning, noise emission and manufacturing processes (e.g. shotpeening) are also closely linked to the phenomenon of jetting \citep{krefting2004high,blake2015}.  The bottleneck in development and comprehension of these techniques as well as naturally occurring phenomenon remains the limited understanding of the interaction amongst cavitation bubble and its surrounding medium. \\

It is well known that the jets formed during the collapse of bubbles in the presence of a nearby wall are directed towards the wall and that the characteristics of these jets can be described well using the scaling laws given by \citet{popinet2002bubble} and \citet{supponen2017collapse}. However, the investigations on the collapse of a bubble initially attached to a wall are reported less often. \citet{naude1961mechanism} discuss the collapse of bubble initially in contact with wall by writing the higher order perturbation solutions to potential flow model. They also show the collapse dynamics of such bubbles experimentally by producing bubbles with electric spark in the liquid close to wall. \citet{hupfeld2020dynamics} present experimental results for laser induced bubbles using ultra-high speed photography in the high capillary number regimes. They study both expansion and collapse phase in several liquids to understand the effect of viscosity on the growth, collapse and bubble shapes including liquid microlayer and contact line evolution. Both \citet{naude1961mechanism} and \citet{hupfeld2020dynamics} show that during the collapse the jet is directed towards the solid whereas \citet{li2018transient} show that the jet can be directed in opposite direction when a collapsing bubble is in contact with a spherical solid particle. Similar problem is studied numerically by \citet{lechner2020jet}, who also include the effect of bubble expansion and the liquid microlayer that results in to different nature of bubble collapse and jetting. This problem has also been briefly discussed in numerical studies of \citet{lauer2012numerical} and \citet{koukouvinis2016compressible} who observe the change in jetting behaviour as a function of bubble shape but they do not show jets opposite to the wall. \citet{reuter2017} speculate the appearance of a jet parallel to the wall during the secondary collapse and consequently the generation of vortex ring, although no direct experimental observation is reported.\\

The initiation and consequences of this change of behaviour has not be understood and the hydrodynamic description of the same is not available.
In this article we bridge this gap and describe the factors controlling the jetting direction using comprehensive theoretical, numerical and experimental evidences. We clarify that the change in the jetting direction is an implication of the singularity in the pressure gradient obtained from the potential flow model. This singularity only appears when the effective contact angle at the instant before the collapse is larger than 90 degrees. We show that at short instants after beginning of collapse phase and for sufficiently large Reynolds and Weber numbers, the potential flow solution accurately represents the results of DNS (Direct Numerical Simulation). Despite the fact that viscosity instantaneously regularise the solution near the singularity and predicts finite velocities, the effect of singularity is still present outside the boundary layer and it explains the changes in the nature of the jet generated. In addition, we show that there is a direct link between the bubble response at short times and the formation of vortex ring that eventually leads to surprising long range interaction effects between the bubble and a free surface (see supplementary video) that have not been reported before. These results provide a significant advance in the characterisation of the interaction mechanisms between a bubble and its surrounding medium and may be a reasoning behind unexplained delayed effects in underwater explosions observed by \citet{kedrinskii1978surface} and tsunami generation by underwater volcano eruptions \citep{paris2015source}.

\section{Problem setup}
\begin{figure}
\centering
\includegraphics{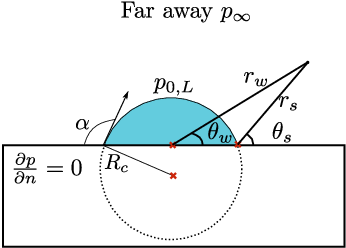}
\caption{Problem setup and different coordinate systems used in this article i.e. $(r_w,\theta_w)$ and $(r_s,\theta_s)$}
\label{fig:setup}
\end{figure}

We focus on the classical three-dimensional Rayleigh collapse problem for a bubble that has a spherical cap shape (figure \ref{fig:setup}). 
This choice is motivated by the fact that for a spherical cap the bubble geometry is defined by only two parameters, the bubble radius and the contact angle, which simplifies the discussion of the results compared to more complex shapes (e.g. ellipsoids) that could be focus of future investigations.
We restrict ourselves to an axisymmetric configuration where the initial bubble pressure $p_{g,0}$ is uniform inside the bubble and the liquid, initially at rest, is at a higher pressure far from the bubble $p_\infty > p_{g,0}$. This problem represents the solution of a bubble in equilibrium state subjected to impulsive action caused by a sudden increase in far-field pressure, and is also a good approximation for the dynamic response of a low pressure bubble after the expansion stage when it reaches its maximum radius and the liquid velocity approaches to zero.\\

The dynamics of this system is governed by the mass, momentum and energy conservation laws, which for the $i$th component in a multi-component setup is given as
\begin{subeqnarray}
        \frac{\partial \rho_i}{\partial t} + \nabla \cdot (\rho_i {\mathbf u}) & = & 0, \label{eq:dnsrho}\\
        \frac{\partial \rho_i {\mathbf u}}{\partial t} + \nabla \cdot (\rho_i {\mathbf u}{\mathbf u}) & = & -\nabla p_i + \nabla \cdot {\mathsfbi \tau}_i, \label{eq:dnsmom}\\
        \frac{\partial (\rho_i E_i)}{\partial t} + \nabla \cdot (\rho_i E_i \mathbf{u}) & = & - \nabla \cdot ({\mathbf u}p_i) + \nabla \cdot ({\mathsfbi \tau}_i {\mathbf u}). \label{eq:dnsenergy}
\end{subeqnarray}
As usual, $\mathbf{u}$ represents the velocity vector field, $\rho_i$ is density, $p_i$ is the pressure field and $\tau_i =  \mu_i (\nabla {\mathbf u} + (\nabla {\mathbf u})^{\rm T})$ is the viscous stress tensor for each component. The total energy $\rho_i E_i = \rho_i e_i + \frac{1}{2} \rho_i {\mathbf u} \cdot {\mathbf u}$, is the sum of internal energy per unit volume $\rho_i e_i$ and kinetic energy $\frac{1}{2} \rho_i {\mathbf u} \cdot {\mathbf u}$ per unit volume. Finally the equation of state (EOS) is written as $\rho_i e_i = \frac{p_i + \Gamma_i \Pi_i}{\Gamma_i - 1}$ where $\Gamma$ and $\Pi$ are adopted from \citet{johnsen2006implementation} in such a way that liquid is a slightly compressible and the bubble contents obeys an ideal gas EOS. The system of equations is closed by adding an advection equation for the interface position which is represented by the volume of fluid method \citep{tryggvason2011direct}. After the advection, the interface is reconstructed using a piecewise linear interface construction. The surface-tension appears implicitly while averaging the equation \ref{eq:dnsrho} in the mixed cells and it is added as the continum surface force  using the method of \citet{brackbill1992continuum} also reviewed recently by \citet{popinet2018numerical}. All the intricate details of the method are given in \citet{fuster2018}\\

In the DNS, a non-slip boundary condition is imposed on the velocity field at the wall, while a standard free flow condition is applied in all the other boundaries. Note that although a non-slip boundary condition imposes zero velocity at the boundary, the velocity at which the interface is advected in the cell adjacent to the wall is different from zero. This effectively acts as an implicit slip length for the contact line motion given as $\lambda = \Delta x_{min}/2$, where $\Delta x_{min}$ is the smallest grid size. This influence of this parameter will be discussed in appendix \ref{appA}. The static contact angle model proposed by \citet{afkhami2008height} is used to implement contact angle boundary condition which although is not very accurate for prediction the contact line motion but is a convenient tool that gives fair insight of the growth of boundary layer and range of influence of viscosity. More sophisticated  models for contact line motion could be used but a significant change in the conclusions of this work is not expected.\\
 
The initial conditions are defined by imposing the fluid to be initially at rest ($\boldsymbol{u}=0$). In this case the divergence of the momentum equation gives a Laplace equation for the initial liquid pressure while assuming the liquid as an incompressible substance,
\begin{equation}
    \Delta p_l = 0.
    \label{eq:laplace}
\end{equation}
This equation is readily solved if we assume that the gas pressure is initially uniform such that the boundary conditions for pressure (as shown in figure \ref{fig:setup}) are the following: at the interface the liquid pressure is 
\begin{equation}
p_{0,L}=p_{g,0} - \frac{2\sigma}{R_c},
\label{eq:laplaceInt}
\end{equation}
where $\sigma$ is the surface tension coefficient and $R_c$ is the radius of curvature; at the far--away boundaries the pressure is $p_\infty$ and at the wall the normal pressure gradient is null due to impermeability condition. The initial condition for total energy is then calculated from the EOS using the initial pressure field.\\
 
We finally introduce relevant dimensionless quantities using the radius of curvature as characteristic length, the liquid density and the characteristic velocity $U_c= \left({p_\infty}/{\rho_l}\right)^{1/2}$ obtained by assuming that at the instant of maximum radius the liquid pressure along the surface is negligible compared to the far--field pressure. A simple dimensional analysis shows that the solution of the problem depends on the density and viscosity ratios, the Reynolds number, $\Rey = \frac{\rho_l U_c R_c}{\mu_l}$, the Weber number $\Web = \frac{\rho_l U_c^2 R_c}{\sigma}$ and the contact angle $(\alpha)$. 
The characteristic velocity is constant and equal to 10 m/s for the low pressure bubbles collapsing in water under the action of the atmospheric pressure, therefore both, the Reynolds and Weber number, are determined only by the radius of curvature of the bubble. 
In the particular experiments considered in this study the radius of curvature remains to be of the order 1 cm obtaining characteristic values of the Weber and Reynolds number of the order of $\Web \sim \textit{O}(10^4)$ and $\Rey \sim \textit{O}(10^5)$. The influence of finite values of Reynolds is briefly discussed using Direct Numerical Simulations.

\begin{figure}
\centering
\includegraphics[]{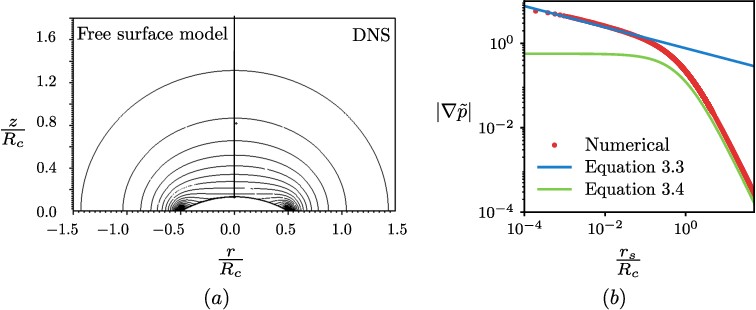}
\caption{$(a)$ Isocontours of the magnitude of acceleration for bubble with $\alpha = \alpha=3\pi/4$: The isocontours in the left half are obtained from free surface model and those in the right half are obtained from DNS. $(b)$ Non-dimensional acceleration magnitude along the wall obtained for bubble with $\alpha=3\pi/4$ using the free surface model.}
\label{fig:contours}
\end{figure}

\section{Short time dynamics}

At short times, we can simplify the system of equations if we consider the liquid as an incompressible substance and the interface as a free surface, i.e. we neglect inertial and viscous effects inside the bubble. This assumption is reasonable given that $\mu_g/\mu_l \ll 1$ and $\rho_g/\rho_l \ll 1$ (in experiments the density ratio is of the order of $10^{-5}$ when the bubble pressure equals the vapour pressure). The velocity at short times remains a small quantity, thus we can neglect the convective terms as compared to temporal derivatives of velocity and spatial pressure gradients \citep{batchelor2000}. This hypothesis remains true when time scales under consideration are smaller than the advection time scales i.e. $\frac{t_s U_c}{R_c}\ll 1$.
Under these assumptions the solution of the velocity is obtained from the integration of the linearised momentum equation as
\begin{equation}
{\mathbf u}(t_s) =  - \frac{1}{\rho} \int_0^{t_s} \nabla p + \frac{1}{\rho} \int_0^{t_s} \nabla \cdot {\mathsfbi \tau}.
\label{eq:momsmalltimes}
\end{equation}
It is classical for this linear system to decompose the velocity field in to a potential part (${\mathbf u}_\phi$) and a viscous correction (${\mathbf u}_\nu$) i.e. ${\mathbf u} =  {\mathbf u}_{\phi} + {\mathbf u}_{\nu}$. The potential flow theory gives ${\mathbf u}_\phi$ whereas the viscous correction requires complex analysis \citep{saffman1995vortex,batchelor2000,pope2001turbulent}. For the short times and large enough Reynolds numbers, the viscous contributions to velocity are relevant only in the small region of thickness $\delta \sim \textit{O} (\sqrt{\nu t})$ near the solid boundary. Thus in such a situation, we expect the solution to be governed by the potential part and the viscous part is only a correction in a thin region close to the wall. We discuss these aspects next.

\subsection{Potential flow solution}
The potential flow solution is obtained by solving the inviscid part of the linearised momentum equation
\begin{equation}
    \frac{\partial {\mathbf u}_{\phi}}{\partial t} = -\frac{1}{\rho} \nabla p.
    \label{eq:Euler2}
\end{equation}
  The magnitude of pressure gradient field (or  acceleration field) at short infinitesimal times govern the dynamics of bubble collapse at a finite time.  To obtain the acceleration we introduce the dimensionless pressure $\tilde{p}=\frac{p-p_{0,L}}{p_\infty - p_{0,L}}$. Note that in the linear problem the interface does not move,
therefore surface tension effects at short times only introduce a correction on the scaling prefactor $(p_\infty-p_{0,L})$ upon which the solution depends. The divergence of equation \ref{eq:Euler2} leads to a Laplace equation ($\nabla^2 \tilde{p} = 0$) which can be solved with appropriate boundary conditions for $\tilde{p}$. In particular we impose $\tilde{p}=1$ far away from the bubble, $\tilde{p}=0$ at the bubble interface and $\frac{d\tilde{p}}{dn} = 0$ at the solid boundary. The solution of this equation depends only on the geometry of the bubble
which is fully described as a function of the contact angle ($\alpha$). This simplified representation of the fluid fields at short times is referred as `free surface model' in this article.\\
 
In figure \ref{fig:contours}a we start presenting the isocontours of the  magnitude of the non-dimensional acceleration ($|\nabla \tilde{p}|$) field obtained for a spherical cap bubble with contact angle $\alpha= \frac{3}{4}\pi$ while figure \ref{fig:contours}b shows the variation of same along the wall in the liquid phase with respect to the distance from the contact line obtained from free surface model. The isocontours in the left half are obtained numerically from the free surface model, while on the right half are obtained from the DNS of the Navier--Stokes equations (using setup shown in figure \ref{fig:boundarylayer}a) accounting for the presence of a gas with non-zero density ($\rho_l/\rho_g = 10^{-3}$) and a finite value of viscosity ($\mu_l/\mu_g = 10^{-2}$) at $\frac{t U_c}{R_c} = 2.07 \cdot 10^{-6}$. Good agreement between the two models supports the argument that equation \ref{eq:laplace} is a good representation of the fluid dynamics at very short times. We clearly distinguish two separate regions: the far-field where the pressure gradient contours are hemispherical caps centred at the axis of symmetry, and the near field where the contours for pressure gradient are intricate and diverge on approaching the triple contact point.
Next, we characterise each of these regions:

\subsubsection{Near field}

\begin{figure}
  \centering
   \includegraphics[]{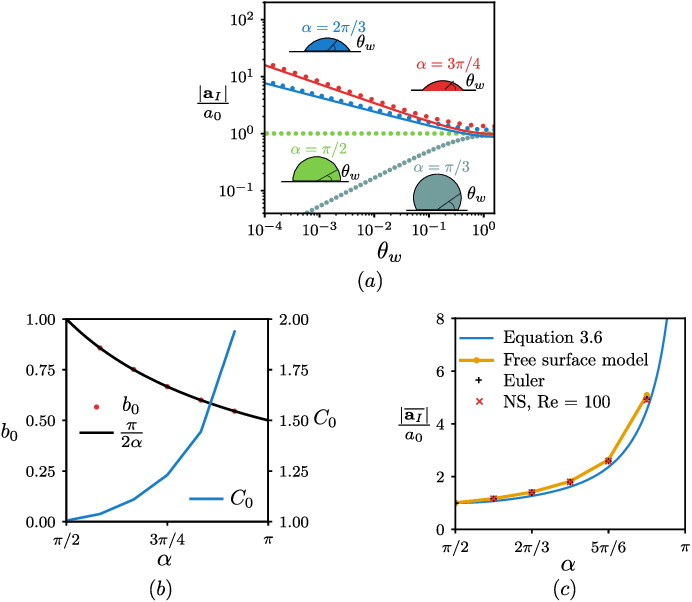}
  \caption{ $(a)$ Non-dimensional acceleration magnitude, ${\vert {\mathbf a}_I\vert }/a_0$ along the interface parametrised using the angle $\theta_w$ (measured in the counter-clockwise direction from the point of contact of wall and the axis of symmetry) for  $\alpha = \frac{\upi}{3},\frac{\pi}{2},\frac{2}{3}\pi,\frac{3}{4}\pi$. Dots represent the numerical solution from free surface model and the thick lines are predictions using equation \ref{eq:derivativenear} evaluated at the interface. $(b)$ Exponent $b_0$ and coefficient $C_0$ obtained from fitting the pressure gradient obtained from the free surface model along the wall   
  for different $\alpha$. $(c)$ Non-dimensional averaged interface  acceleration magnitude as a function of the contact angle using different methods: analytical expression given by equation \ref{eq:avgacc} (blue line), the solution from the free surface model (yellow line),  the DNS solution of the Euler equations (black crosses), and the DNS solution of the Navier--Stokes solver for $\Rey=100$ (red crosses).}
\label{fig:interface}
\end{figure}

The numerical solution of the interface acceleration at the initial time
is shown in figure \ref{fig:interface}a, where we use the angle with respect to the solid
wall $\theta_w$ to parametrise the interface position. The origin, $\theta_w = 0$, corresponds to the point of contact between the interface and the solid wall and 
$\theta_w=\pi/2$ is the intersection of the interface with the axis of symmetry.
Figure \ref{fig:interface}a depicts the local non-dimensional acceleration magnitude at bubble interface $\vert {\mathbf a} \vert/a_0$ for four representative cases with $\alpha=\upi/3, \upi/2, 2\upi/3,3\upi/4$. 
The value of the reference acceleration is chosen as $a_0 = \frac{p_\infty - p_{0,L}}{\rho R_c}$, which is the acceleration magnitude for a spherical bubble with radius $R_c$ and a known pressure difference. 
As expected, in the case of $\alpha = \upi/2$, the numerical solution recovers the Rayleigh-Plesset solution and the non-dimensional acceleration is uniform and equal to one all over the interface. When $\alpha < \upi/2$ the interface acceleration (thus velocity) tends to zero at the contact point, and therefore it seizes to move even in the potential flow problem with a slip wall. For $\alpha > \upi/2$, the appearance of the singularity is evident as the acceleration diverges
as we approach the $\theta_w\to 0$ limit.
To interpret this result one must keep in mind that 
the general solution of the Laplace equation in spherical coordinates can be obtained by separation of variables that leads to an infinite series. This series sometimes diverges at the edge where Dirichlet and Neumann boundary conditions meet due to the appearance of a singularity \citep{dauge2006elliptic}. This singularity was reviewed extensively in the past \citep{landau,blum1982finite,steger1990corner,deegan1997capillary,li2000} but the studies where this singularity appears in the bubble dynamics problems are rare. The asymptotic solutions close to the point where homogeneous Dirichlet and Neumann boundary conditions meet can be alternatively expressed by using the general solution of the Laplace equation \citep{li2000,dauge2006elliptic,yosibash2011circular} and takes the form of
\begin{equation}
    \tilde{p}_s= \sum_k^\infty C_k \tilde{r}_s^{b_k} \cos\left(b_k \theta_s \right),
    \label{eq:nearfieldfit}  
\end{equation}
where $\tilde{r}_s=r_s/R_c$ is the non-dimensional distance from the contact line and $b_k=\frac{\upi}{\alpha} (k+1/2)$. Taking the derivative with respect to the normal of the interface we readily find the interface acceleration magnitude near the contact line as
\begin{equation}
    \vert {\mathbf a}_I \vert = \frac{1}{\rho} \left| \frac{\partial p}{\partial n} \right| \approx a_0 C_0 b_0 \tilde{r}_s^{b_0 - 1},
    \label{eq:derivativenear}
\end{equation}
This expression exhibits a singularity at the contact point ($r_s \to 0$) when $b_0 = \frac{\upi}{2\alpha} < 1$ (or $\alpha > \upi/2$) implying that the acceleration at the triple contact point is infinite. In these conditions the first term in the expansion is the leading order term that eventually dominates the solution in the region $r_s \le R_c$.
This can also be clearly seen in figure \ref{fig:contours}b where the fitting curve shown with solid line is obtained using the first term of the series solution (i.e. parameters $C_0$ and $b_0$ in equation \ref{eq:derivativenear}).
We repeat the fitting procedure for various $\alpha$ to verify that the numerical values of $b_0$ match well with theoretical predictions and find $C_0$ which is a constant of order one that slightly increases with $\alpha$ (figure \ref{fig:interface}b).\\

Focusing on the characterisation of the regimes where the singularity appears, figure \ref{fig:interface}a
 clearly shows that the numerical solution (dots) is well described by equation \ref{eq:derivativenear} evaluated at the interface (solid lines), showing excellent agreement close to the contact point, i.e. small values of $\theta_w$. Near the axis of symmetry (e.g. $\theta_w\to \upi/2$ and $\tilde{r}_s \approx 1$), the errors are visible and the first term in the series does not suffice to describe accurately the acceleration field.

\subsubsection{Far field solution}
The far field flow created by the bubble corresponds to a punctual sink sitting at the intersection between the wall and the axis of symmetry. The integration of the momentum equation in the radial coordinate provides the magnitude of pressure gradient generated by a punctual sink at an arbitrary distance from the sink as
\begin{equation}
\left| \frac{d\tilde{p}_{\rm{far}}}{d\tilde{r}_w} \right| = \frac{\vert \overline{\mathbf a}_I \vert}{a_0} \frac{(1 + \cos(\alpha))}{\tilde{r}^2_w}, 
\label{eq:farfieldfit}
\end{equation} 
where  $\vert \overline{\mathbf a}_I \vert$ is the magnitude of averaged acceleration along the interface, which determines the strength of the punctual sink. Figures \ref{fig:contours}b clearly shows that this equation captures well the decay of the pressure gradient with the distance $\tilde{r}_w$ far from the interface.
Because the first term of the series i.e. equation \ref{eq:derivativenear} predicts the interface acceleration reasonably well, we obtain an estimation of the averaged acceleration magnitude for bubbles with $\alpha > \upi/2$  as
\begin{equation}
 \vert \overline{\mathbf a}_I \vert = \left| \frac{1}{S_b} \int \frac{-1}{\rho_l}\frac{\partial p}{\partial n} dS_b \right| = a_0 C_0 b_0 G(\alpha),
\label{eq:avgacc}
\end{equation}
where $S_b$ stands for the bubble interface surface and
    $G(\alpha)$ is a geometrical factor that can be numerically computed assuming that the first term in the series is indeed the leading order term along the entire bubble interface.  
Figure \ref{fig:interface}c shows that the averaged nondimensional interface acceleration magnitude obtained using this model, which compares well with the numerical results of the free surface model and the solution of the both viscous and inviscid Navier-Stokes equations. The results of the simplified model are obtained using the value of $C_0$ numerically computed and reported in figure \ref{fig:interface}b.
For $\alpha > \upi/2$, $\vert \overline{\mathbf a} \vert$ increases with $\alpha$. The three  proposed models agree well for the angles tested implying that the free surface model as well as the simplified expression given in equation \ref{eq:avgacc} capture well the averaged bubble response at short times for sufficiently large Reynolds numbers.


\subsection{Viscous correction} \label{sec:visc}
The potential flow solution is only formally exact at the initial time $t = 0$ when the interface is at rest. As soon as the interface is set in to motion a thin boundary layer immediately develops regularising the flow close to the contact point. We use the results from the DNS of the Navier--Stokes equations at short times to investigate the influence of viscosity on the dynamic response of the bubbles.\\
   
  \begin{figure}
    \centering
    
    \includegraphics[scale = 1]{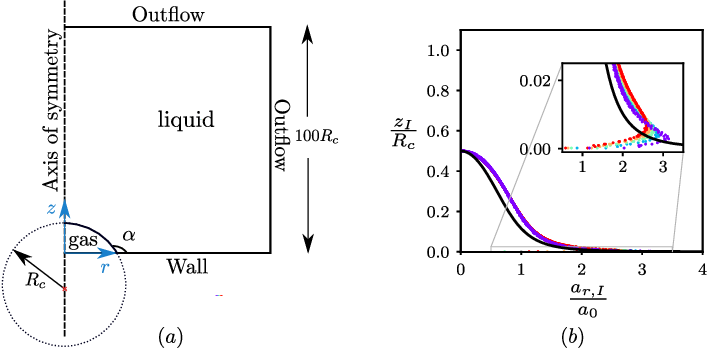}
      \caption{ $(a)$ The numerical setup for DNS $(b)$ Results from the DNS for $\alpha = 2 \upi /3$ and $\Rey = 100$. Time averaged interface acceleration in direction parallel to the wall, $a_{r,I} = u_{r,I}/t$,  as a function of the distance from the wall at five different times (in colour where $t U_c/R_{c,0}\in [7.27 \cdot 10^{-5},3.6 \cdot 10^{-4}]$). For reference we include the potential flow solution given by equation \ref{eq:derivativenear} evaluated at the interface (solid black line). The inset represents a zoom in to the viscous boundary layer generated close to the wall. } 
\label{fig:boundarylayer}
\end{figure}

We solve for the full set of equations \ref{eq:dnsrho} where, for the sake of simplicity, we neglect surface tension effects. The equations are solved using the All-Mach implementation \citep{popinet2015quadtree,fuster2018}, where the one fluid approach is used to solve 
the modelling equations. This solver has been already used successfully to investigate some aspects on the interaction of an oscillating bubble
and a rigid tube by \citet{fan2020optimal} and interactions of collapsing bubbles with free surface by \cite{saade2021crown}. The setup for numerical simulations is shown in figure \ref{fig:boundarylayer}a, where a spherical cap-shaped bubble is initialised in a square domain of size 100 times the initial radius of curvature of this bubble. We use such large domains so that the pressure waves reflected from the boundaries do not affect the solution, moreover grid is coarsened away from the bubble to diffuse these waves. The simulations are axisymmetric about left boundary and the bubble is initially in contact with the bottom boundary where a non-slip boundary conditions is applied. The faraway boundaries are given constant far-field pressure conditions and homogeneous Neumann condition for velocity. The grid is refined progressively near the interface to a minimum grid size of $\Delta x_{min}/R_c = 0.00038$. The maximum time step in the simulation is governed by acoustic CFL restriction, which in our case is 0.4. The initial pressure field inside the domain is given from Equation \ref{eq:laplace}, velocity field is given as zero and total energy is given from EOS.\\

In figure \ref{fig:boundarylayer}b we characterise the bubble motion at short non-dimensional time $t U_c/R_{c}$ by showing the evolution of the time averaged interface acceleration magnitude parallel to the wall, defined as  $a_{r}(t) = u_r(t)/t$,  as a function of the normal distance from the wall. The results shown with coloured dots are obtained from DNS for $\alpha = 2\upi/3$ and $\Rey = 100$ where colour scale represents the non-dimensional time varying between $7.27 \cdot 10^{-5}$ and $3.6\cdot 10^{-4}$.
The inset figure shows the clear development of a boundary layer very close to the wall, the thickness of this layer defined by the height for which the velocity is maximum. Outside this region the free surface potential flow solution (equation \ref{eq:derivativenear}), shown with a solid black line,  predicts the interface velocity obtained from DNS relatively well, therefore the potential flow solution is a good approximation of the collapse process outside the viscous boundary layer. Inside the viscous boundary layer, the interface acceleration is sensitive to the slip length imposed (see Appendix \ref{appA}) and also to the movement of the contact line. In this region, the solution of free surface model at $t=0$ cannot be extrapolated to predict the flow field near the contact line at short times. Remarkably, although the contact line motion is grid dependent, the average acceleration and the maximum jet velocity is shown to be independent of the slip length imposed (see Appendix \ref{appA}). This fact together with slight dependence of the average acceleration with $\Rey$ (figure \ref{fig:interface}c) confirms that for large enough Reynolds numbers, the bubble dynamic response is mainly governed by the potential flow solution and not by the particular contact line motion model used.  In these conditions, the free surface potential flow model at $t=0$ is a useful tool to predict the average dynamics of the interface at short times. \\
  
\begin{figure}
\centering
\includegraphics[]{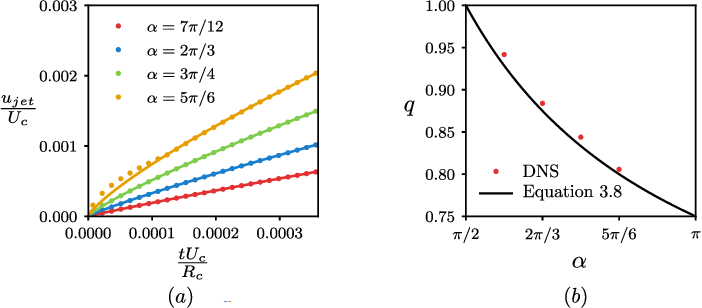}\\
  \caption{ $(a)$ Evolution of jet velocity for different contact angles from (dots) DNS and (solid lines) fitting using equation \ref{eq:powerlaw}. $(b)$ Exponent $q$ from DNS fitting and predictions of equation \ref{eq:qvariation}.} 
\label{fig:boundarylayerfit}
\end{figure}

In order to characterise the jet formation process we extract the maximum interface velocity magnitude from DNS data, which we name as jet velocity ${u}_{jet} $.
This velocity is fitted using a power law function of the form (figure \ref{fig:boundarylayerfit}a) 
\begin{equation}
 \frac{ u_{jet}}{U_c} = A \left( \frac{t U_c}{R_c} \right)^q,
\label{eq:powerlaw}
\end{equation}
where $q$ is a coefficient that becomes smaller than one when the acceleration is singular
at $t=0$.
A theoretical estimation of coefficient $q$ accounting for viscous effects can be obtained from the singular solution described in the previous section as following: The jet velocity after short time $(t_s)$ can be approximated from the initial acceleration evaluated at the interface at height equal to thickness of boundary layer  $z=\delta(t_s)$. The $\delta$ is given approximately by the solution of the Stokes problem for the flow near the flat plate that is impulsively started from rest therefore, we can impose that  $\delta(t_s) \approx C_\delta \sqrt{\nu t_s}$ where $C_\delta$ is a constant of order unity. In this case the jet velocity velocity is estimated as $$u_{jet} (t_s) = a_{r,I} \left(t=0, z_I=\delta(t_s)\right) \; t_s$$ where we assume that the interface quickly decelerates as soon as it enters inside the viscous boundary layer.
Under these assumptions, the coefficient $q$ is readily obtained as
 \begin{eqnarray}
 && q=\frac{1}{2} + \frac{\upi}{4\alpha}. \label{eq:qvariation}
 \end{eqnarray}
 
As shown in Figures \ref{fig:boundarylayerfit}b this model is capable of capturing the evolution of the jet velocity at short times found from DNS simulations accounting for viscous effects when  $\alpha > \upi/2$. The exponent $q$ (figure \ref{fig:boundarylayerfit}b) decreases as $\alpha$ increases because the acceleration gradient in the normal direction increases with $\alpha$ as seen in figure \ref{fig:interface}a. Note that the DNS results confirm that  $q < 1$ for $\alpha > \upi/2$ and therefore the jet acceleration is singular at $t=0$ even in the presence of viscous effects. \\

\begin{figure}
  \centering
  \includegraphics[]{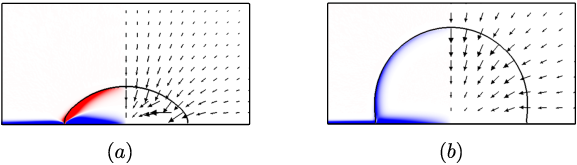}
  \caption{
    The vorticity is shown in colour map and velocity vectors obtained from DNS after short time short time ($\frac{t U_c}{R_c} = 0.1$) for $\rm{Re} = 100$ for $(a)$ $\alpha = 2 \upi/3$ and $(b)$ $\alpha = 5 \upi/12$.}
\label{fig:vorticity}
\end{figure}

Another interesting quantity to describe the overall interface dynamics is the vorticity generated during the collapse. At short times, vorticity is concentrated in a sheet vortex at the wall and at the interface. By assuming zero stress condition at the interface for $t = 0$, the only non-zero component of the vorticity can be written as $$\omega_\phi =  - 2 \frac{\partial {u}_r}{\partial \theta}.$$
It follows from the potential flow solution (figure \ref{fig:interface}a) that the sign of the acceleration gradient changes depending upon initial contact angle i.e. if it is greater or less than $\upi/2$, reverting the sign of the vortex sheet  generated at the interface. This behaviour is indeed observed from the numerical simulations as shown in figure \ref{fig:vorticity} (a and b), where the strength of the vortex sheet is plotted with saturated colour maps. The change in dominant colour represents the opposite sign of vorticity at interface between the two cases. 
 
\section{Long time dynamics}\label{sec:jetting}

\subsection{High Reynolds regime}
\begin{figure}
  \centering
    \includegraphics[scale = 0.8]{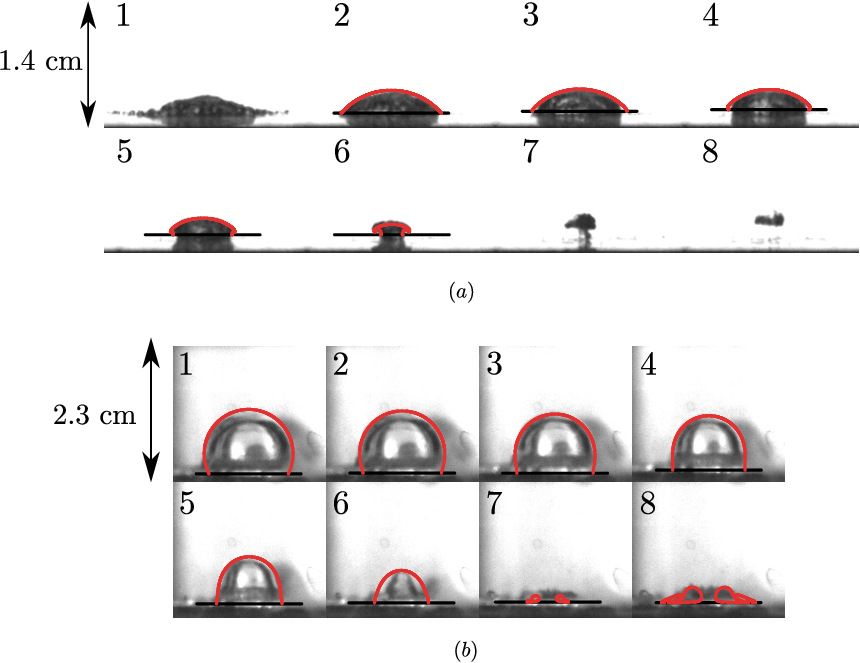}
  \caption{
  The snapshots of bubble shape are shown for the two representative cases, the numerical bubble shapes shown with red curves are overlaid at same times and scaled to same length as in experiments. $(a)$ Case where $\alpha > \upi/2$, each snapshots are taken every $0.1$ ms $(b)$ Case where $\alpha < \upi/2$, each snapshots are taken every $0.125$ ms.
  }
\label{fig:DNSvortfield}
\end{figure}

After describing the short time dynamics, in this section we investigate the long time consequences of this singular response using the experiments and DNS. The results for two representative cases $\alpha > \pi/2$ and $\alpha < \pi/2$ shown in figure \ref{fig:DNSvortfield} $(a)$ and $(b)$ are discussed now.\\

\subsubsection{Dynamics of bubbles with an effective contact angle larger than $\pi/2$}

First we discuss the case where the singularity is present (e.g. $\alpha > \pi/2$). The generation of bubbles with $\alpha > \pi/2$  can be investigated using the setup described briefly in appendix \ref{appB} and references \citep{bourguille2017shock,tahan2020evolution}. The laser is focused at the bottom of liquid tank where a water drop is attached. The cavitation in the water drop induces shock wave in aluminium plate that leads to the appearance of multiple bubbles in the water tank that interact to form a flat bubble with shape similar to spherical cap at maximum volume  (figure \ref{fig:DNSvortfield} a). The snapshots are taken at every $0.1$ms, the interface from the second snapshot is extracted and fitted with an approximate spherical cap which gives $R_c = 7.56\times 10^{-3} m$ and $\alpha = 0.727 \pi$.\\

We reproduce the experimental conditions numerically using the numerical setup described in section \ref{sec:visc}. The minimum mesh size of $\Delta x = 60 \mu m$, the far field pressure $p_\infty$ is 1 atm and the pressure inside bubble is set to a low value of 0.1 atm which is chosen by hit and trial method to match the numerical collapse time with the collapse time in experiments. We consider both surface tension and viscous effects in the numerical simulations with $\Rey \approx 75600$ and $\Web \approx 10500$. The numerically obtained bubble shapes plotted with red contours and scaled to bubble size in snapshot 2, these are subsequently overlaid on the experimental snapshots after same times. A very good agreement is seen between the numerical and the experimental bubble shapes, small differences subsist because of
the simplification of the bubble shape during the expansion, the influence of gravity, mass transfer and other effects that are not considered in the numerical simulations.\\

\begin{figure}
\centering
  \includegraphics[scale = 0.75]{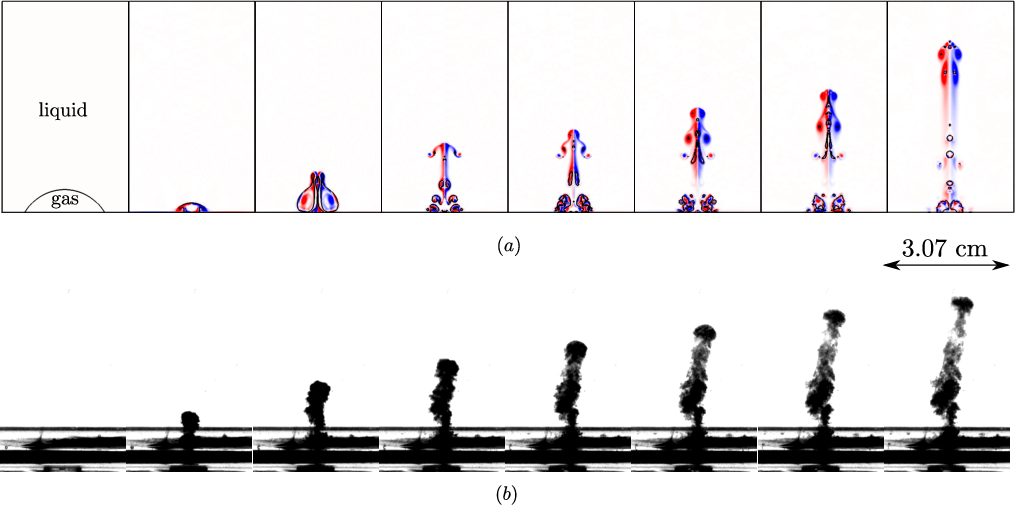}\\
  \caption{$(a)$ Bubble interface in black curve and the vorticity field in the colour map as obtained from DNS is shown at (consecutively from left to right) $\frac{t U_c}{R_c}= 0,0.19,0.3,0.81,1.00,1.45,1.95,4.90$. An the annular jet parallel to wall reaches the axis resulting in to the jet directed away from the wall which later generates the vortex ring. The results are plotted for $\alpha = 2\upi/3$ and $\Rey = \infty$ $\Web = \infty$ $(b)$ The visualisation of liquid flow field obtained by adding dye at the bottom of tank resulting from collapse of bubble is shown at every 2.5ms.}
\label{fig:vortexring}
\end{figure}

At the beginning of collapse phase the highest interface velocity is developed at edge of the viscous boundary layer (see figure \ref{fig:boundarylayer}b) that leads to the appearance of an annular jet which is generated parallel to the wall and a mushroom-like shape of the interface contour (see figure \ref{fig:DNSvortfield}a and supplementary movie 1). These results are consistent with previous numerical works of \citet{lauer2012numerical} and \citet{koukouvinis2016compressible}. Remarkably, when the collapse is strong enough and the jet reaches the axis of symmetry, a stagnation point appears there and a secondary upward jet normal to the wall is generated.
The re-entrant jet observed for $\alpha > \pi/2$ is not conventional in cavitation and generates vortex ring structures similar to those observed by \citet{reuter2017} which are persistent in nature and can travel large distances in comparison to bubble size. The generation of this vortex ring is illustrated numerically in figure \ref{fig:vortexring}a where the colour maps show the vorticity field. A clear re-entrant annular jet is observed, followed by a mushroom like structure and the bubble detaches from the wall consequently a jet in the upwards direction which eventually leads to formation of vortex ring. In the experiments we visualise this by adding the dye in the bottom of the tank (figure \ref{fig:vortexring}b) during the collapse of a flat bubble. This vortex ring can travel to long distance and induce unusual long range effects like free surface waves and jetting (see supplementary movie 3).
\\

\subsubsection{Dynamics of bubbles with an effective contact angle smaller than $\pi/2$}

The dynamics of bubbles with contact angles smaller than 90 degrees (figure \ref{fig:DNSvortfield} b) can be obtained using a classical experiment where a laser is focused directly in to the liquid very close to the wall. The bubble shape from snapshot 2 in figure \ref{fig:DNSvortfield}b is approximated with a spherical cap that gives 
$R_c = 7.63\times 10^{-3} m$ and $\alpha = 0.389 \pi$. The interface contours obtained numerically are shown with the red curves scaled to size of the bubble size in first snapshot. In this case, the interface acceleration is minimum at the contact line and maximum at the tip of the spherical cap leading to a conventional high--speed liquid jet directed towards the wall (see figure \ref{fig:DNSvortfield}b and supplementary movie 2). This jet developed towards the wall impinges in to the solid surface causes cavitation damage. Similar dynamics have been described in several of previous studies \citep{naude1961mechanism,gonzalez2021}.

\subsection{Finite Reynolds number effects}

\citet{popinet2002bubble} showed that for the small bubbles collapsing in the vicinity of a rigid wall, the jets can be suppressed due to the viscous effects. In this section, we use the DNS to quantitatively predict the range of size for which 
the observations reported in this manuscript apply. We take an ideal case of an air bubble with $p_{b,0} = 0.1$ atm and constant contact angle equal to 120 degrees collapsing in water at an ambient far field pressure.
In this situation, the characteristic velocity of the process stays constant and equal to 10 m/s and the initial radius of curvature is then left as the unique parameter controlling the values of the Reynolds and Weber number. \\

\begin{figure}
    \centering
    \includegraphics{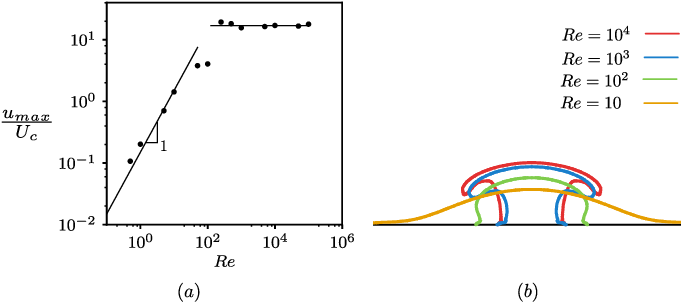}
    \caption{$(a)$ Peak of the non-dimensional velocity as a function of Reynolds number
    for an air bubble at $p_0=0.1$ atm collapsing in water at atmospheric pressure. $(b)$ Interface contours as a function of the Reynolds number at the instant of minimum radius.}
    \label{fig:umaxvsoh}
\end{figure}

In figure \ref{fig:umaxvsoh}a we show the maximum  non-dimensional velocity reached during the collapse as a function of the Reynolds number. 
For Reynolds numbers above a given critical value $\Rey > \Rey_c= 100$ the influence of Reynolds on the peak velocities is only marginal. However, when the Reynolds is below this critical value, the jet velocity drops dramatically further decaying with decreasing $\Rey$ . This sudden change on the peak velocities is controlled by the appearance of jetting, which is not visible for $\Rey \leq 100$
(see figure \ref{fig:umaxvsoh}b).
For a low pressure air bubble collapsing in water at atmospheric pressure these results reveal that the observations reported (reversed re-entrant jet) are applicable for bubbles larger than $R_c > 10 \mu$m. We must still emphasise that the results reported here are limited to spherical cap bubbles. In reality other kind of behaviour can be observed depending upon the mechanism by which the bubbles are generated and the bubble shape reached at the instant of maximum expansion. If the bubble shape is not a spherical cap or the microlayer is not able to drain during the bubble collapse or the bubble size is comparable to pits, it can lead to a variety of other collapse and jetting dynamics discussed by \citet{lechner2020jet}, \citet{reuter2021supersonic} and \cite{trummler2020near} respectively.

\section{Conclusions}
In this work, we show that the impulsive potential flow theory
can be used to discuss the influence of the bubble shape on the dynamic
response of collapsing bubbles for sufficiently
large Reynolds and Weber numbers.  As an example, we present the results obtained for the collapse of the spherical cap bubbles showing that the effective contact angle at the instant of maximum expansion controls the interface acceleration at the beginning of the collapse phase and the jetting direction observed in Direct Numerical Simulations and experiments. 
 When $\alpha > 90$ degrees the potential flow solution 
at small times shows the appearance of a singularity which causes extremely high accelerations close to the contact point and a change in vorticity direction
with respect to the $\alpha \le 90$ degrees case.
 In the former case, an unconventional jetting mechanism is observed and shown to be responsible for the appearance of a vortex dipole travelling in the direction opposite to the wall.
 The nature of interaction between the bubble and the surrounding medium
is then strongly influenced by the bubble shape at the instant before collapse, 
appearing as a critical parameter if one wants to control or model the physical phenomena triggered by the collapse of bubbles attached to a wall.
 \\

The findings of this work could be used to explain and control various processes triggered by the collapse of bubbles. Some examples include mixing and transport processes, the development of treatment technologies based on the bubble--tissue interactions including drug delivery and high intensity ultrasound techniques \citep{prentice2005membrane,ohl2006sonoporation,maxwell2011cavitation} and the interactions between a collapsing bubble and the free surface \citep{kedrinskii1978surface}. Another exciting phenomenon is generation of surface waves from the bubble collapse in deep water. The phenomenon described in this manuscript may be eventually related to the appearance of Tsunamis from underwater volcanic eruptions where the mechanisms linking both processes are not fully understood yet \cite{paris2015source}. Finally the fundamental understanding of interaction of collapsing bubble with the surrounding media should improve the design and development of various technologies based on the physical, mechanical and chemical effects induced by the collapse of bubbles. The relation of the singularity of potential flow to the well known force singularity during the motion of contact line described by \cite{huh1971hydrodynamic} remains another open research problem where the solutions reported in this work may inspire new ideas.


\backsection[Acknowledgements]{Authors want to acknowledge Julien Le Clanche, research engineer at ENSTA Bretagne for his kind support.}

\backsection[Funding]{This research is supported by European Union (EU) under the MSCA-ITN grant agreement number 813766, under the project named ultrasound cavitation in soft matter (UCOM). Part of this work was part of the PROBALCAV program supported by The French National Research Agency (ANR) and cofunded by DGA (French Minisitry of Defense Procurment Agency) under reference Projet ANR-21-ASM1-0005 PROBALCAV.}

\backsection[Declaration of interests]{The authors report no conflict of interest.}



\backsection[Author contributions]{D.F., M.A. and S.Z. designed the research program. M.S. conducted the simulations and analyzed data. M.S. and D.F. did the theoretical calculus. E.T. and M.A. performed the experiments. M.S, D.F, M.A., E.T. and S.Z discussed the results. M.S. and D.F. wrote the paper.}

\appendix

\section{Convergence study}\label{appA}

 \begin{figure}
  \centering
  \includegraphics[]{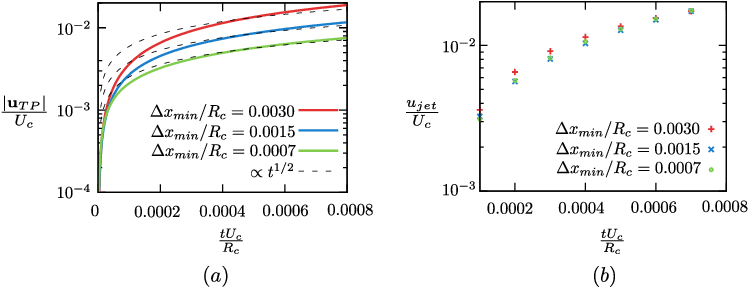}\\
  \caption{The grid convergence of viscous solution $(a)$ The evolution of velocity of contact line is plotted for different grid-size. $(b)$ The evolution of jet velocity is plotted for various grid size.}
\label{fig:convergence}
\end{figure}

 The convergence properties of the contact line velocity and jet velocity 
 are discussed in this section. The velocity of the contact line (figure \ref{fig:convergence}a where $\vert {\mathbf u}_{TP}\vert$ is contact line velocity) changes as the mesh is refined since slip length changes implicitly ($\lambda = \Delta x_{min}/2$). It is also evident from the same plot that in very short times the well known scaling for the thickness of the boundary layer ($t^{1/2}$) is not recovered as the boundary layer is not well resolved. At slightly longer times the boundary layer growth recovers the expected scaling law. Despite of the sensitivity of the contact line motion to the slip length imposed, the peak interface velocity (jet velocity $u_{jet}$) plotted in Figure \ref{fig:convergence}b reveal that the numerical results are converged and therefore independent of the slip length.

\section{Experimental method}
\label{appB}

\begin{figure}
  \centering
  \includegraphics{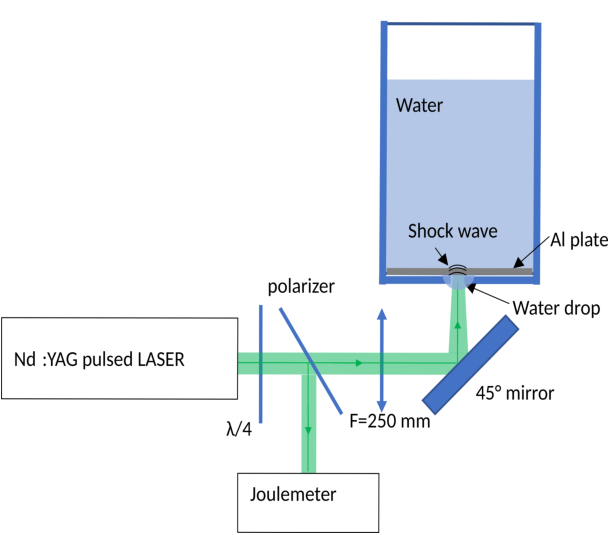}
  \caption{Experimental setup used to create flat bubbles.}
\label{fig:experimental}
\end{figure}

The experiments are performed with pulsed Nd:YAG laser (model quanta ray pro 350-10) that provides maximum of 3.5J of energy with Gaussian distribution in time having half maximum at 9.2 ms. This laser beam passes through a quarter wave plate and polariser for modulating the energy and then focused using a planoconvex lens of focal length 250mm that creates a focal spot of 4mm in diameter on the aluminium plate. For bubbles shown with $\alpha > \upi/2$ the laser is focused on the water droplet attached to bottom of an aluminium plate (1 mm thick) kept at bottom of a water tank and for the bubbles with $\alpha < \upi/2$ the laser if focused on the top of this aluminium plate kept at the bottom of same tank (see figure \ref{fig:experimental}). The laser energy is set to 50\% of maximum in case where $\alpha < \upi/2$ and we observe the cavitation phenomenon in the water tank using high speed camera.

\bibliographystyle{jfm}
\bibliography{biblio}

\begin{thebibliography}{55}
\expandafter\ifx\csname natexlab\endcsname\relax\def\natexlab#1{#1}\fi
\def\au#1{#1} \def\ed#1{#1} \def\yr#1{#1}\def\at#1{#1}\def\jt#1{\textit{#1}}
  \def\bt#1{#1}\def\bvol#1{\textbf{#1}} \def\vol#1{#1} \def\pg#1{#1}
  \def\publ#1{#1}\def\arxiv#1{#1}\def\org#1{#1}\def\st#1{\textit{#1}}

\bibitem[Afkhami \& Bussmann(2008)]{afkhami2008height}
{\sc \au{Afkhami, S} \& \au{Bussmann, M}} \yr{2008}  \at{Height functions for
  applying contact angles to 2d vof simulations}.  \jt{International journal
  for numerical methods in fluids}  \bvol{57}~(4),  \pg{453--472}.

\bibitem[Atchley \& Prosperetti(1989)]{atchley1989crevice}
{\sc \au{Atchley, Anthony~A} \& \au{Prosperetti, Andrea}} \yr{1989}  \at{The
  crevice model of bubble nucleation}.  \jt{The Journal of the Acoustical
  Society of America}  \bvol{86}~(3),  \pg{1065--1084}.

\bibitem[Batchelor(2000)]{batchelor2000}
{\sc \au{Batchelor, George~k}} \yr{2000} {\em An introduction to fluid
  dynamics\/}.  \publ{Cambridge university press}.

\bibitem[Bhat \& Gogate(2021)]{bhat2021cavitation}
{\sc \au{Bhat, Akash~P} \& \au{Gogate, Parag~R}} \yr{2021}
  \at{Cavitation-based pre-treatment of wastewater and waste sludge for
  improvement in the performance of biological processes: A review}.
  \jt{Journal of Environmental Chemical Engineering}  \bvol{9}~(2),
  \pg{104743}.

\bibitem[Blake {\em et~al.\/}(2015)Blake, Leppinen \& Wang]{blake2015}
{\sc \au{Blake, John~R}, \au{Leppinen, David~M} \& \au{Wang, Qianxi}} \yr{2015}
   \at{Cavitation and bubble dynamics: the kelvin impulse and its
  applications}.  \jt{Interface focus}  \bvol{5}~(5),  \pg{20150017}.

\bibitem[Blum \& Dobrowolski(1982)]{blum1982finite}
{\sc \au{Blum, H} \& \au{Dobrowolski, M}} \yr{1982}  \at{On finite element
  methods for elliptic equations on domains with corners}.  \jt{Computing}
  \bvol{28}~(1),  \pg{53--63}.

\bibitem[Borkent {\em et~al.\/}(2009)Borkent, Gekle, Prosperetti \&
  Lohse]{borkent2009nucleation}
{\sc \au{Borkent, Bram~M}, \au{Gekle, Stephan}, \au{Prosperetti, Andrea} \&
  \au{Lohse, Detlef}} \yr{2009}  \at{Nucleation threshold and deactivation
  mechanisms of nanoscopic cavitation nuclei}.  \jt{Physics of fluids}
  \bvol{21}~(10),  \pg{102003}.

\bibitem[Bourguille {\em et~al.\/}(2017)Bourguille, Bergamasco, Tahan, Fuster
  \& Arrigoni]{bourguille2017shock}
{\sc \au{Bourguille, Judith}, \au{Bergamasco, Luca}, \au{Tahan, Gilles},
  \au{Fuster, Daniel} \& \au{Arrigoni, Michel}} \yr{2017} Shock propagation
  effects in multilayer assembly including a liquid phase.  \bt{In {\em Key
  engineering materials\/}}, ,  \vol{vol. 755},  \pg{pp. 181--189}. Trans Tech
  Publ.

\bibitem[Brackbill {\em et~al.\/}(1992)Brackbill, Kothe \&
  Zemach]{brackbill1992continuum}
{\sc \au{Brackbill, Jeremiah~U}, \au{Kothe, Douglas~B} \& \au{Zemach, Charles}}
  \yr{1992}  \at{A continuum method for modeling surface tension}.  \jt{Journal
  of computational physics}  \bvol{100}~(2),  \pg{335--354}.

\bibitem[Brenner {\em et~al.\/}(2002)Brenner, Hilgenfeldt \&
  Lohse]{brenner2002single}
{\sc \au{Brenner, Michael~P}, \au{Hilgenfeldt, Sascha} \& \au{Lohse, Detlef}}
  \yr{2002}  \at{Single-bubble sonoluminescence}.  \jt{Reviews of modern
  physics}  \bvol{74}~(2),  \pg{425}.

\bibitem[Cole \& Weller(1948)]{cole1948}
{\sc \au{Cole, Robert~H} \& \au{Weller, Royal}} \yr{1948}  \at{Underwater
  explosions}.  \jt{Physics Today}  \bvol{1}~(6),  \pg{35}.

\bibitem[Crum(1979)]{crum1979tensile}
{\sc \au{Crum, Lawrence~A}} \yr{1979}  \at{Tensile strength of water}.
  \jt{Nature}  \bvol{278}~(5700),  \pg{148--149}.

\bibitem[Dauge(2006)]{dauge2006elliptic}
{\sc \au{Dauge, Monique}} \yr{2006} {\em Elliptic boundary value problems on
  corner domains: smoothness and asymptotics of solutions\/}, ,  \vol{vol.
  1341}.  \publ{Springer}.

\bibitem[Deegan {\em et~al.\/}(1997)Deegan, Bakajin, Dupont, Huber, Nagel \&
  Witten]{deegan1997capillary}
{\sc \au{Deegan, Robert~D}, \au{Bakajin, Olgica}, \au{Dupont, Todd~F},
  \au{Huber, Greb}, \au{Nagel, Sidney~R} \& \au{Witten, Thomas~A}} \yr{1997}
  \at{Capillary flow as the cause of ring stains from dried liquid drops}.
  \jt{Nature}  \bvol{389}~(6653),  \pg{827--829}.

\bibitem[Fan {\em et~al.\/}(2020)Fan, Li \& Fuster]{fan2020optimal}
{\sc \au{Fan, Yuzhe}, \au{Li, Haisen} \& \au{Fuster, Daniel}} \yr{2020}
  \at{Optimal subharmonic emission of stable bubble oscillations in a tube}.
  \jt{Physical Review E}  \bvol{102}~(1),  \pg{013105}.

\bibitem[Fuster(2019)]{fuster2019review}
{\sc \au{Fuster, Daniel}} \yr{2019}  \at{A review of models for bubble clusters
  in cavitating flows}.  \jt{Flow, Turbulence and Combustion}  \bvol{102}~(3),
  \pg{497--536}.

\bibitem[Fuster \& Popinet(2018)]{fuster2018}
{\sc \au{Fuster, Daniel} \& \au{Popinet, St{\'e}phane}} \yr{2018}  \at{An
  all-mach method for the simulation of bubble dynamics problems in the
  presence of surface tension}.  \jt{Journal of Computational Physics}
  \bvol{374},  \pg{752--768}.

\bibitem[Gonzalez-Avila {\em et~al.\/}(2021)Gonzalez-Avila, Denner \&
  Ohl]{gonzalez2021}
{\sc \au{Gonzalez-Avila, Silvestre~Roberto}, \au{Denner, Fabian} \& \au{Ohl,
  Claus-Dieter}} \yr{2021}  \at{The acoustic pressure generated by the
  cavitation bubble expansion and collapse near a rigid wall}.  \jt{Physics of
  Fluids}  \bvol{33}~(3),  \pg{032118}.

\bibitem[Hou {\em et~al.\/}(2021)Hou, Wang, Yan, Wang \& An]{hou2021study}
{\sc \au{Hou, Zhaoning}, \au{Wang, Lin}, \au{Yan, Xiaona}, \au{Wang, Zhanwei}
  \& \au{An, Libei}} \yr{2021}  \at{Study on characteristics of the cavitation
  bubble dynamics of lithium bromide aqueous solution with ultrasonic
  interaction}.  \jt{Journal of Building Engineering}  \bvol{44},  \pg{102424}.

\bibitem[Huh \& Scriven(1971)]{huh1971hydrodynamic}
{\sc \au{Huh, Chun} \& \au{Scriven, Laurence~E}} \yr{1971}  \at{Hydrodynamic
  model of steady movement of a solid/liquid/fluid contact line}.  \jt{Journal
  of colloid and interface science}  \bvol{35}~(1),  \pg{85--101}.

\bibitem[Hupfeld {\em et~al.\/}(2020)Hupfeld, Laurens, Merabia, Barcikowski,
  G{\"o}kce \& Amans]{hupfeld2020dynamics}
{\sc \au{Hupfeld, Tim}, \au{Laurens, Ga{\'e}tan}, \au{Merabia, Samy},
  \au{Barcikowski, Stephan}, \au{G{\"o}kce, Bilal} \& \au{Amans, David}}
  \yr{2020}  \at{Dynamics of laser-induced cavitation bubbles at a
  solid--liquid interface in high viscosity and high capillary number regimes}.
   \jt{Journal of Applied Physics}  \bvol{127}~(4),  \pg{044306}.

\bibitem[Johnsen \& Colonius(2006)]{johnsen2006implementation}
{\sc \au{Johnsen, Eric} \& \au{Colonius, Tim}} \yr{2006}  \at{Implementation of
  weno schemes in compressible multicomponent flow problems}.  \jt{Journal of
  Computational Physics}  \bvol{219}~(2),  \pg{715--732}.

\bibitem[Kedrinskii(1978)]{kedrinskii1978surface}
{\sc \au{Kedrinskii, VK}} \yr{1978}  \at{Surface effects from an underwater
  explosion}.  \jt{Journal of Applied Mechanics and Technical Physics}
  \bvol{19}~(4),  \pg{474--491}.

\bibitem[Koukouvinis {\em et~al.\/}(2016)Koukouvinis, Gavaises, Georgoulas \&
  Marengo]{koukouvinis2016compressible}
{\sc \au{Koukouvinis, Phoevos}, \au{Gavaises, Manolis}, \au{Georgoulas,
  Anastasios} \& \au{Marengo, Marco}} \yr{2016}  \at{Compressible simulations
  of bubble dynamics with central-upwind schemes}.  \jt{International Journal
  of Computational Fluid Dynamics}  \bvol{30}~(2),  \pg{129--140}.

\bibitem[Krefting {\em et~al.\/}(2004)Krefting, Mettin \&
  Lauterborn]{krefting2004high}
{\sc \au{Krefting, D}, \au{Mettin, Robert} \& \au{Lauterborn, Werner}}
  \yr{2004}  \at{High-speed observation of acoustic cavitation erosion in
  multibubble systems}.  \jt{Ultrasonics Sonochemistry}  \bvol{11}~(3-4),
  \pg{119--123}.

\bibitem[Landau \& Lifshitz(1959)]{landau}
{\sc \au{Landau, Lev~D} \& \au{Lifshitz, Evgueni}} \yr{1959} {\em Fluid
  Mechanics\/}, ,  \vol{vol.~6}.  \publ{Pergamon Press}.

\bibitem[Lauer {\em et~al.\/}(2012)Lauer, Hu, Hickel \&
  Adams]{lauer2012numerical}
{\sc \au{Lauer, E}, \au{Hu, XY}, \au{Hickel, Stefan} \& \au{Adams,
  Nikolaus~Andreas}} \yr{2012}  \at{Numerical modelling and investigation of
  symmetric and asymmetric cavitation bubble dynamics}.  \jt{Computers \&
  Fluids}  \bvol{69},  \pg{1--19}.

\bibitem[Lechner {\em et~al.\/}(2020)Lechner, Lauterborn, Koch \&
  Mettin]{lechner2020jet}
{\sc \au{Lechner, Christiane}, \au{Lauterborn, Werner}, \au{Koch, Max} \&
  \au{Mettin, Robert}} \yr{2020}  \at{Jet formation from bubbles near a solid
  boundary in a compressible liquid: Numerical study of distance dependence}.
  \jt{Physical Review Fluids}  \bvol{5}~(9),  \pg{093604}.

\bibitem[Li {\em et~al.\/}(2018)Li, Zhang, Wang \& Han]{li2018transient}
{\sc \au{Li, Shuai}, \au{Zhang, A-Man}, \au{Wang, Shiping} \& \au{Han, Rui}}
  \yr{2018}  \at{Transient interaction between a particle and an attached
  bubble with an application to cavitation in silt-laden flow}.  \jt{Physics of
  Fluids}  \bvol{30}~(8),  \pg{082111}.

\bibitem[Li \& Lu(2000)]{li2000}
{\sc \au{Li, Zi-Cai} \& \au{Lu, Tzon-Tzer}} \yr{2000}  \at{Singularities and
  treatments of elliptic boundary value problems}.  \jt{Mathematical and
  Computer Modelling}  \bvol{31}~(8-9),  \pg{97--145}.

\bibitem[Lohse {\em et~al.\/}(2001)Lohse, Schmitz \& Versluis]{lohse2001}
{\sc \au{Lohse, Detlef}, \au{Schmitz, Barbara} \& \au{Versluis, Michel}}
  \yr{2001}  \at{Snapping shrimp make flashing bubbles}.  \jt{Nature}
  \bvol{413}~(6855),  \pg{477--478}.

\bibitem[Lyons {\em et~al.\/}(2019)Lyons, Haney, Fee, Wech \&
  Waythomas]{lyons2019infrasound}
{\sc \au{Lyons, John~J}, \au{Haney, Matthew~M}, \au{Fee, David}, \au{Wech,
  Aaron~G} \& \au{Waythomas, Christopher~F}} \yr{2019}  \at{Infrasound from
  giant bubbles during explosive submarine eruptions}.  \jt{Nature Geoscience}
  \bvol{12}~(11),  \pg{952--958}.

\bibitem[Maxwell {\em et~al.\/}(2013)Maxwell, Cain, Hall, Fowlkes \&
  Xu]{maxwell2013}
{\sc \au{Maxwell, Adam~D}, \au{Cain, Charles~A}, \au{Hall, Timothy~L},
  \au{Fowlkes, J~Brian} \& \au{Xu, Zhen}} \yr{2013}  \at{Probability of
  cavitation for single ultrasound pulses applied to tissues and
  tissue-mimicking materials}.  \jt{Ultrasound in medicine \& biology}
  \bvol{39}~(3),  \pg{449--465}.

\bibitem[Maxwell {\em et~al.\/}(2011)Maxwell, Wang, Cain, Fowlkes, Sapozhnikov,
  Bailey \& Xu]{maxwell2011cavitation}
{\sc \au{Maxwell, Adam~D}, \au{Wang, Tzu-Yin}, \au{Cain, Charles~A},
  \au{Fowlkes, J~Brian}, \au{Sapozhnikov, Oleg~A}, \au{Bailey, Michael~R} \&
  \au{Xu, Zhen}} \yr{2011}  \at{Cavitation clouds created by shock scattering
  from bubbles during histotripsy}.  \jt{The Journal of the Acoustical Society
  of America}  \bvol{130}~(4),  \pg{1888--1898}.

\bibitem[Naud{\'e} \& Ellis(1961)]{naude1961mechanism}
{\sc \au{Naud{\'e}, Charl~F.} \& \au{Ellis, Albert~T.}} \yr{1961}  \at{{On the
  Mechanism of Cavitation Damage by Nonhemispherical Cavities Collapsing in
  Contact With a Solid Boundary}}.  \jt{Journal of Basic Engineering}
  \bvol{83}~(4),  \pg{648--656},  \arxiv{arXiv:
  https://asmedigitalcollection.asme.org/fluidsengineering/article-pdf/83/4/648/5486537/648\_1.pdf}.

\bibitem[Ohl {\em et~al.\/}(2006{\natexlab{{\em a\/}}})Ohl, Arora, Dijkink,
  Janve \& Lohse]{ohl2006}
{\sc \au{Ohl, Claus-Dieter}, \au{Arora, Manish}, \au{Dijkink, Rory}, \au{Janve,
  Vaibhav} \& \au{Lohse, Detlef}} \yr{2006{\natexlab{{\em a\/}}}}  \at{Surface
  cleaning from laser-induced cavitation bubbles}.  \jt{Applied physics
  letters}  \bvol{89}~(7),  \pg{074102}.

\bibitem[Ohl {\em et~al.\/}(2006{\natexlab{{\em b\/}}})Ohl, Arora, Ikink,
  De~Jong, Versluis, Delius \& Lohse]{ohl2006sonoporation}
{\sc \au{Ohl, Claus-Dieter}, \au{Arora, Manish}, \au{Ikink, Roy}, \au{De~Jong,
  Nico}, \au{Versluis, Michel}, \au{Delius, Michael} \& \au{Lohse, Detlef}}
  \yr{2006{\natexlab{{\em b\/}}}}  \at{Sonoporation from jetting cavitation
  bubbles}.  \jt{Biophysical journal}  \bvol{91}~(11),  \pg{4285--4295}.

\bibitem[Paris(2015)]{paris2015source}
{\sc \au{Paris, Rapha{\"e}l}} \yr{2015}  \at{Source mechanisms of volcanic
  tsunamis}.  \jt{Philosophical Transactions of the Royal Society A:
  Mathematical, Physical and Engineering Sciences}  \bvol{373}~(2053),
  \pg{20140380}.

\bibitem[Pishchalnikov {\em et~al.\/}(2019)Pishchalnikov, Behnke-Parks,
  Schmidmayer, Maeda, Colonius, Kenny \& Laser]{pishchalnikov2019high}
{\sc \au{Pishchalnikov, Yuri~A}, \au{Behnke-Parks, William~M}, \au{Schmidmayer,
  Kevin}, \au{Maeda, Kazuki}, \au{Colonius, Tim}, \au{Kenny, Thomas~W} \&
  \au{Laser, Daniel~J}} \yr{2019}  \at{High-speed video microscopy and
  numerical modeling of bubble dynamics near a surface of urinary stone}.
  \jt{The Journal of the Acoustical Society of America}  \bvol{146}~(1),
  \pg{516--531}.

\bibitem[Pope(2001)]{pope2001turbulent}
{\sc \au{Pope, Stephen~B}} \yr{2001} Turbulent flows.

\bibitem[Popinet(2015)]{popinet2015quadtree}
{\sc \au{Popinet, St{\'e}phane}} \yr{2015}  \at{A quadtree-adaptive multigrid
  solver for the serre--green--naghdi equations}.  \jt{Journal of Computational
  Physics}  \bvol{302},  \pg{336--358}.

\bibitem[Popinet(2018)]{popinet2018numerical}
{\sc \au{Popinet, St{\'e}phane}} \yr{2018}  \at{Numerical models of surface
  tension}.  \jt{Annual Review of Fluid Mechanics}  \bvol{50},  \pg{49--75}.

\bibitem[Popinet \& Zaleski(2002)]{popinet2002bubble}
{\sc \au{Popinet, Stephane} \& \au{Zaleski, Stephane}} \yr{2002}  \at{Bubble
  collapse near a solid boundary: a numerical study of the influence of
  viscosity}.  \jt{Journal of fluid mechanics}  \bvol{464},  \pg{137--163}.

\bibitem[Prado {\em et~al.\/}(2022)Prado, Antunes, Rocha,
  S{\'a}nchez-Mu{\~n}oz, Barbosa, Ter{\'a}n-Hilares, Cruz-Santos, Arruda,
  da~Silva \& Santos]{prado2022review}
{\sc \au{Prado, CA}, \au{Antunes, FAF}, \au{Rocha, TM},
  \au{S{\'a}nchez-Mu{\~n}oz, S}, \au{Barbosa, FG}, \au{Ter{\'a}n-Hilares, R},
  \au{Cruz-Santos, MM}, \au{Arruda, GL}, \au{da~Silva, SS} \& \au{Santos, JC}}
  \yr{2022}  \at{A review on recent developments in hydrodynamic cavitation and
  advanced oxidative processes for pretreatment of lignocellulosic materials}.
  \jt{Bioresource Technology}  \bvol{345},  \pg{126458}.

\bibitem[Prentice {\em et~al.\/}(2005)Prentice, Cuschieri, Dholakia, Prausnitz
  \& Campbell]{prentice2005membrane}
{\sc \au{Prentice, Paul}, \au{Cuschieri, Alfred}, \au{Dholakia, Kishan},
  \au{Prausnitz, Mark} \& \au{Campbell, Paul}} \yr{2005}  \at{Membrane
  disruption by optically controlled microbubble cavitation}.  \jt{Nature
  physics}  \bvol{1}~(2),  \pg{107--110}.

\bibitem[Reuter {\em et~al.\/}(2017)Reuter, Gonzalez-Avila, Mettin \&
  Ohl]{reuter2017}
{\sc \au{Reuter, Fabian}, \au{Gonzalez-Avila, Silvestre~Roberto}, \au{Mettin,
  Robert} \& \au{Ohl, Claus-Dieter}} \yr{2017}  \at{Flow fields and vortex
  dynamics of bubbles collapsing near a solid boundary}.  \jt{Physical Review
  Fluids}  \bvol{2}~(6),  \pg{064202}.

\bibitem[Reuter \& Ohl(2021)]{reuter2021supersonic}
{\sc \au{Reuter, Fabian} \& \au{Ohl, Claus-Dieter}} \yr{2021}  \at{Supersonic
  needle-jet generation with single cavitation bubbles}.  \jt{Applied Physics
  Letters}  \bvol{118}~(13),  \pg{134103}.

\bibitem[Saade {\em et~al.\/}(2021)Saade, Jalaal, Prosperetti \&
  Lohse]{saade2021crown}
{\sc \au{Saade, Youssef}, \au{Jalaal, Maziyar}, \au{Prosperetti, Andrea} \&
  \au{Lohse, Detlef}} \yr{2021}  \at{Crown formation from a cavitating bubble
  close to a free surface}.  \jt{Journal of Fluid Mechanics}  \bvol{926}.

\bibitem[Saffman(1995)]{saffman1995vortex}
{\sc \au{Saffman, Philip~G}} \yr{1995} {\em Vortex dynamics\/}.
  \publ{Cambridge university press}.

\bibitem[Steger(1990)]{steger1990corner}
{\sc \au{Steger, Karl}} \yr{1990}  \at{Corner singularities of solutions of the
  potential equation in three dimensions}.  \bt{In {\em Mathematical Modelling
  and Simulation of Electrical Circuits and Semiconductor Devices\/}},  \pg{pp.
  283--297}.  \publ{Springer}.

\bibitem[Supponen(2017)]{supponen2017collapse}
{\sc \au{Supponen, Outi}} \yr{2017}  \bt{Collapse phenomena of deformed
  cavitation bubbles}. {\em Tech. Rep.\/}.  \org{EPFL}.

\bibitem[Tahan {\em et~al.\/}(2020)Tahan, Arrigoni, Bidaud, Videau \&
  Th{\'e}venet]{tahan2020evolution}
{\sc \au{Tahan, Gilles}, \au{Arrigoni, Michel}, \au{Bidaud, Pierre},
  \au{Videau, Laurent} \& \au{Th{\'e}venet, David}} \yr{2020}  \at{Evolution of
  failure pattern by laser induced shockwave within an adhesive bond}.
  \jt{Optics \& Laser Technology}  \bvol{129},  \pg{106224}.

\bibitem[Trummler {\em et~al.\/}(2020)Trummler, Bryngelson, Schmidmayer,
  Schmidt, Colonius \& Adams]{trummler2020near}
{\sc \au{Trummler, Theresa}, \au{Bryngelson, Spencer~H}, \au{Schmidmayer,
  Kevin}, \au{Schmidt, Steffen~J}, \au{Colonius, Tim} \& \au{Adams,
  Nikolaus~A}} \yr{2020}  \at{Near-surface dynamics of a gas bubble collapsing
  above a crevice}.  \jt{Journal of Fluid Mechanics}  \bvol{899}.

\bibitem[Tryggvason {\em et~al.\/}(2011)Tryggvason, Scardovelli \&
  Zaleski]{tryggvason2011direct}
{\sc \au{Tryggvason, Gr{\'e}tar}, \au{Scardovelli, Ruben} \& \au{Zaleski,
  St{\'e}phane}} \yr{2011} {\em Direct numerical simulations of gas--liquid
  multiphase flows\/}.  \publ{Cambridge university press}.

\bibitem[Yosibash {\em et~al.\/}(2011)Yosibash, Shannon, Dauge \&
  Costabel]{yosibash2011circular}
{\sc \au{Yosibash, Zohar}, \au{Shannon, Samuel}, \au{Dauge, Monique} \&
  \au{Costabel, Martin}} \yr{2011}  \at{Circular edge singularities for the
  laplace equation and the elasticity system in 3-d domains}.
  \jt{International journal of fracture}  \bvol{168}~(1),  \pg{31--52}.

\end{thebibliography}

\end{document}